\documentclass[prd,twocolumn,aps,showpacs]{revtex4}

\newcommand{\be}{\begin{equation}}
\newcommand{\ee}{\end{equation}}
\newcommand{\ea}{\end{eqnarray}}
\newcommand{\ba}{\begin{eqnarray}}
\def\p{\partial}
\def\l{\label}

\begin{document}


\title{Open string with a background B-field as the first order mechanics, \\ noncommutativity and soldering formalism}

\author{A. A. Deriglazov$^a$\footnote{\sf E-mail: alexei.deriglazov@ufjf.edu.br}, C. Neves$^b$\footnote{\sf E-mail: cneves@fisica.ufjf.br}, W. Oliveira$^b$\footnote{\sf E-mail: wilson@fisica.ufjf.br}, E. M. C. Abreu$^c$\footnote{\sf E-mail: evertonabreu@ufrrj.br}, C. Wotzasek$^{d}$\footnote{\sf E-mail: clovis@if.ufrj.br} and C. Filgueiras$^e$} 
\affiliation{${}^{a}$Departamento de Matem\'atica,ICE, Universidade Federal de Juiz de Fora,\\
36036-330, Juiz de Fora, MG, Brazil\\
${}^{b}$Departamento de F\'{\i}sica,ICE, Universidade Federal de Juiz de Fora,\\
36036-330, Juiz de Fora, MG, Brazil\\
${}^{c}$Departamento de F\'{\i}sica, Universidade Federal Rural do Rio de Janeiro\\
BR 465-07, 23851-180, Serop\'edica, Rio de Janeiro, Brazil\\
${}^{d}$Instituto de F\'\i sica, Universidade Federal do Rio de
Janeiro, 21945, Rio de Janeiro, Brazil\\
${}^{e}$Departamento de F\'{\i}sica, CCEN, Universidade Federal da Para\'{\i}ba, Cidade
Universit\'aria, 58051-970 Jo\~ao Pessoa, PB, Brazil\\
\bigskip
\today}


\begin{abstract}
\noindent To study noncommutativity properties of the open string with constant
B-field we construct a mechanical action which reproduces classical
dynamics of the string sector under consideration. It allows one to
apply the Dirac quantization procedure for constrained systems in
a direct and unambiguous way. The mechanical action turns out to be the
first order system without taking the strong field limit
$B\longrightarrow\infty$. In particular, it is true for zero mode
of the string
coordinate which means that the noncommutativity is intrinsic property
of this mechanical system. We describe the arbitrariness in the relation existent 
between the mechanical and the string variables and show that 
noncommutativity of the  string variables on the boundary can be removed. 
It is in correspondence with the result
of Seiberg and Witten on relation among noncommutative and ordinary
Yang-Mills theories.  The recently developed soldering formalism helps us to establish
a connection between the original open string action and the Polyakov action.
\end{abstract}

\maketitle


\section{Introduction}
 
Noncommutative Yang-Mills theory arises in a definite limit of string
theory \cite{cds}. It has been extracted by Seiberg and Witten  \cite{sw} 
starting from the open string in the presence of a B-field [3-9], with the action for the
corresponding sector being \cite{sw}
\begin{eqnarray}\label{1}
S=-\frac{1}{4\pi\alpha^\prime}\int d^2\sigma
\left[\partial_ax^i\partial_ax^i+2\pi\alpha^\prime\epsilon^{ab}
\partial_ax^i\partial_bx^jB_{ij}\right]\,\,,
\end{eqnarray}
where we adopted Euclidean metrics both on the world sheet and in spacetime.
An extremum of the action is supplied by
\begin{eqnarray}\label{2}
(\partial^2_\tau-\partial^2_\sigma)x^i = 0,
\end{eqnarray}
\begin{eqnarray}\label{3}
\lbrace \partial_\sigma x^i+2\pi\alpha^\prime\partial_\tau x^jB_j{}^i\rbrace
\mid_{\sigma=0}^{\sigma=\pi} = 0.
\end{eqnarray}
Open string propagator with the boundary conditions (\ref{3}) contains
an antisymmetric matrix \cite{clny,ft}, which gives rise to noncommutativity of
the string coordinate on the boundaries \cite{sw,schomerus}. It was suggested \cite{aas,aas2,aas3,ch}
that the noncommutative geometry can be reproduced in the elementary
framework of constrained systems as a result of the Hamiltonian
quantization of the system (\ref{2}), (\ref{3}), by analogy with the
noncommutativity arising for coordinates of a charged particle in
the lowest Landau level \cite{gj,dj,bs}. To achieve this, there were proposed
rather radical modifications \cite{aas,ch} of the Dirac procedure for
constrained systems. There is some discrepancy among the results obtained
in different approaches, which was discussed in \cite{aas,aas2,aas3,ch,ch2,lee,ko}.  
Let us talk about these issues for a while in order to put our work in a perspective.

In \cite{aas} the authors analyze the connection among the various appearances of noncommutativity
in string theory, $M$-theory and $M$(atrix) model.  The work relates the noncommutativity in $M$(atrix)
compactification to a phenomenon of noncommutativity of space due to mixed boundary conditions in string theory
(see \cite{aas} and references therein).  It was shown that the noncommutativity of the canonical commutation relations of the space 
coordinates of the mixed membrane reflects the zero brane distribution inside the $D$-membrane and therefore is closely related
to the noncommutativity of the $D$-brane  dynamics.  The presence of a Kalb-Ramond field requires the consideration of wrapping the
mixed branes over a torus in order to recover the noncommutativity of the torus of the $M$(atrix) model compactification.  
The $T$-duality helps obtaining the mass spectrum of the open strings compactified on a noncommutative torus.

The generalization for the case of a $D$-string on a torus in a non-zero $B_{\mu\nu}$ background field, i.e., a moving $D$-string which has a non-zero electric field was obtained in \cite{aas2} using a Dirac-Born-Infeld action in a $B$ field background which gives the dynamics of  $D$-strings.  The Hamiltonian formulation helps in the construction of the mass spectrum of the membrane and the in full BPS spectrum.   It was shown that the noncommutativity of brane coordinates come about naturally in the formulation of open strings in the background $B$ field, as well as the noncommutativity of the torus.   The $C^*$-algebra of functions on the noncommutative torus was constructed using the noncommutative open string position operators.

In both papers above it was shown that noncommutativity can be derived within the string theory by embedding branes with
$B_{ij}$ background field on the compactification torus.   In \cite{aas3} the authors used the basic principles of Dirac quantization in order to discuss the noncommutativity of the internal coordinates of branes.  It was shown that the noncommutativity of the coordinates of the open string-brane system is intrinsic, namely, there is no gauge in which the noncommutativity can be removed.  There are alterations in the noncommutativity of the system since the gauge keeps changing, but it never disappear.  And the position of noncommutativity is always on the brane.  In this work \cite{aas3} the mixed boundary conditions were treated as the Dirac constraints and the quantization was obtained through the Dirac method without mode expansions.

On the other hand, in \cite{ch}, the mixed boundary conditions make the canonical quantization of the theory non-trivial.  An inconsistency was obtained when standard commutation relations were imposed.  This inconsistency was removed through the relaxation of the commutativity of the space coordinates of the open strings along the direction of the brane described by mixed boundary conditions.  The analysis was based on a time averaged symplectic form.  The conclusion was that mixed open strings are noncommutative only at the end, where open strings are attached to the brane, and the center of mass coordinates are commuting.   Moreover, the noncommutative end coordinates lead to the conclusion that the $D$-brane world-volume is a noncommutative space.  And the $D$-brane world volume does not need to be a compact space in order to have noncommutativity.  The quantization method used in \cite{ch} was based on the string mode expansion.

In the paper \cite{ch2} the authors obtained the same result as before in \cite{ch} although using the Dirac constrained quantization method.  The constraints are also the boundary conditions of the open string ending on a $D$-brane.  The appearance of an ambiguity during the quantization process justifies the use of a lattice regularization by replacing the interval $[0,\pi]$ by a lattice of $N$ equidistant points with spacing $\epsilon=\pi/N$.  The conclusion is that a $D$-brane in the $B$ field background has a noncommutative world volume in the sense that its world volume theory is the supersymmetric Yang-Mills theory living on a noncommutative space.  The authors proposed that the arbitrariness of the regularization can lead to different results in the literature.
Therefore, trying to make a comparison, we can say that in \cite{aas3} the nontrivial boundary condition in the presence of the $B$-field modify the canonical commutation relations and leads to noncommutativity on $D$-brane world volume.  In \cite{ch2} the authors examine the symplectic form obtained in terms of the mode expansion of the classical solutions.   The detail that characterizes the difference between both works is the fact that the canonical Hamiltonian, instead of the primary Hamiltonian, has been used in the stability condition of the primary constraint, which is different from the usual Dirac quantization formalism.

To understand the dynamics of $D$-brane with a $B$-field, the author in \cite{lee} used the canonical quantization.  With background field $B$ one can obtain a mixed boundary condition, which generates an infinite number of secondary second-class constraints.  These constraints were solved explicitly in \cite{lee} in order to obtain a simple Hamiltonian for the open string in the $D$-brane background.  The Hamiltonian found was a free Hamiltonian for an open string in spacetime.

Finally, in \cite{ko} the Dirac constraint quantization of the brane on the constant antisymmetric backgrounds was reconsidered to obtain consistent noncommutative commutators of the canonical variables including momenta.  The main difference between \cite{aas3,ch2} and \cite{ko} is the choice of the nonvanishing multiplier in the primary Hamiltonian.  The results disclose  no secondary constraints and the single primary constraint itself forms the second class constraint algebra.   The final result show several kinds of interesting commutators and expected noncommutative coordinates.

\bigskip

The aim of our work is to quantize the system (\ref{2}), (\ref{3})
following the standard methods \cite{dirac,gt} without any modifications.
Canonical analysis of this system presents a problem
since the Dirac procedure is initially formulated for the mechanical
system (and then can be generalized for a field with vanishing boundary
conditions). Application of the Dirac formalism to a field with nontrivial
boundary conditions on a compact manifold requires more careful analysis
(see \cite{bh} for the open string with the Neumann boundary conditions).
Consistent treatment of such a system implies necessity to represent
the initial dynamics in terms of a mechanical system. Thus we first
rewrite Eqs. (\ref{2}), (\ref{3}) in the form of equations of motion for
mechanical variables $c_n(\tau), ~ n\in Z,$ and then restore the 
mechanical action
which reproduces this dynamics. After that, the Dirac procedure can be
applied to the mechanical system in a direct and unambiguous way. In
particular, $\delta$-function regularization is not necessary in this
case. The last step is to rewrite the results in terms of the string
coordinate $x^i(\tau, \sigma)$, which gives the Hamiltonian formulation 
associated with the theory (\ref{1})-(\ref{3}).

Some of the results thus obtained are as follows.
The corresponding mechanical system 
turns out to be the first order system without taking the
strong field limit $B\longrightarrow\infty$. 
Thus, as a consequence of the mixed boundary conditions, dynamics of 
the mechanical variables is
governed by equations of the first order in time derivative. In
particular,
it is true for the zero modes $c_0^i(\tau)$ of the string coordinate
$x^i(\tau, \sigma)$. As a consequence, $c_0^i$ are canonically conjugated
to each other: $\{c_0^i, c_0^j\}\ne 0$. It means that the noncommutativity
is intrinsic property of this mechanical system. Brackets for the string
coordinates turn out to be noncommutative in the bulk and on the boundary.
We study freedom in relation between the mechanical and the string
variables as well as freedom in choosing brackets which is always
present in the Hamiltonian formalism. This allows one to discuss to what
extent the noncommutativity of the string coordinates can be avoided.
We show that the embedding coordinates of a D-brane can be made
commutative, of course one needs in this case to transform 
simultaneously the Hamiltonian of the system. 
It is in correspondence with the result of Seiberg and Witten on
equivalence of noncommutative and commutative Yang-Mills fields.

This paper is organized in the following manner: in section 2 we present a solution for the problem
(\ref{2}) and (\ref{3}) with our mechanical variables and consequently the equations of motion.  
In section 3 we apply directly the Dirac algorithm to the action obtained in section 2.
Second class constraints are obtained and consequently we obtain the Dirac brackets for the mechanical
variables and the respective Hamiltonian.  In section 4, with the results obtained in the previous sections, we construct a Hamiltonian formulation for the open string with a $B$-field and for the coordinates on $D$-brane.  In section 5 we discuss the freedom in the definition of the phase space bracket in the Hamiltonian formalism.  In section 6 we accomplish the soldering procedure between different aspects of the open string with a $B$-field.    In the final section we make the conclusions and perspectives.

\section{Open string with B-field in terms of mechanical variables} 

To start with, let us continue $x^i(\sigma), ~ \sigma\in [0, \pi ]$ on the interval 
$[0, 2\pi]$ such that $\tilde x^i(\sigma)=x^i(\sigma)$ on $[0, \pi], ~ 
\partial_\sigma\tilde x\mid_\pi=\vec\partial_\sigma x\mid_\pi, \ldots ,$ 
and $\tilde x^i(2\pi)=\tilde x^i(0), ~ \vec\partial_\sigma\tilde x^i
\mid_{2\pi}=\stackrel{\leftarrow}{\partial}_\sigma
\tilde x^i\mid_0, \ldots ~.$ It can be 
further continued on $\sigma\in(-\infty, \infty)$ as a periodic function 
$\tilde x^i(\sigma +2\pi n)=\tilde x^i(\sigma)$. As a result, any 
solution of the problem (\ref{2}), (\ref{3}) can be presented in the form 

\begin{eqnarray}\label{4}
x^i(\tau, \sigma)=c_0^i(\tau)+\sum_{n=1}^\infty\frac 1n
\left[c_n^i(\tau) \cos n\sigma+c_{-n}^i(\tau)\sin n\sigma\right],\nonumber \\
\mbox{}
\end{eqnarray}

\noindent where $c_n^i(\tau), ~ n\in Z$, are our mechanical variables. Substitution 
in Eqs. (\ref{2}), (\ref{3}) gives the equations of motion 
\begin{eqnarray}\label{5}
\ddot c_0^i=0, \qquad \ddot c_n^i+n^2c_n^i=0, \qquad n\ne 0; 
\end{eqnarray}
\begin{eqnarray}\label{6}
\dot c_0^i=0, \qquad \dot c_n^jB_j{}^i+
\frac{n}{2\pi\alpha^\prime}c_{-n}^i=0, 
\qquad n>0,
\end{eqnarray}
from which one finds further consequence
 
\begin{eqnarray}
\label{6.1}
\dot c^i_{-n}- 2\pi\alpha^\prime nc^j_nB_j{}^i=0.
\end{eqnarray} 
\par
\noindent Consequently, the complete dynamics can be rewritten in the 
equivalent form as 
\begin{eqnarray}\label{7}
\dot c_0^i&=&0, \cr
\dot c_n^jB_j{}^i+\frac{n}{2\pi\alpha^\prime}c_{-n}^i&=&0, \cr
\dot c_{-n}^i-2\pi\alpha^\prime nc_n^jB_j{}^i&=&0, \qquad n>0,
\end{eqnarray}
which consist of equations of the first order only.
Eq. (\ref{7}) follows from the first order action 
\begin{eqnarray}\label{8}
S_f=&&\int d\tau
\left[\frac 12\dot c_0^iB_{ij}c_0^j+\sum_{n=1}^\infty f_n\left(
\pi\alpha^\prime
\dot c_n^iB_{ij}c_n^j-\right.\right. \cr
&&\left.\left.\frac{1}{4\pi\alpha^\prime}\dot c_{-n}^iB^{-1}_{ij}
c_{-n}^j+nc_{-n}^ic_n^i\right)\right],
\end{eqnarray}
where $f_n\ne 0, ~ n>0$ are real numbers. While they can be removed by 
shift of the variables $c_n$, it is convenient to keep them in the 
action. Starting from any particular $S_f$, the variables $c_n$ can be 
taken as the ones which generate the string coordinate (\ref{4}), the 
latter will not depend on $f_n$. At the same time, brackets of $c_n$ 
and the Hamiltonian $H_f$ will depend on $f_n$, so there is appear 
natural arbitrariness in the induced bracket for $x^i(\tau, \sigma)$. 
Choice of some particular form of the bracket implies that the 
corresponding Hamiltonian $H_f$ must be associated with the Lagrangian 
formulation (\ref{1})-(\ref{3}). 

Other possibility is to take as independent the equations (\ref{5}),(\ref{6}) 
with $n>0$. Then the action can be chosen in the second order form 
\begin{eqnarray}\label{700}
S&=&\int d\tau
\left[\frac 12\dot c_0Bc_0+\sum_{n=1}^\infty\frac 12
\dot c_n\dot c_n-\frac{n^2}{2}c_nc_n \right. \nonumber \\
&+& \left.(\dot c_nB+
\frac{n}{2\pi\alpha\prime}c_{-n})\lambda_n\right]\,\,.
\end{eqnarray}
It looks less natural since it involves the Lagrangian multipliers 
$\lambda_n$.

\section{Hamiltonian analysis of the mechanical action}
 
Direct application of the Dirac algorithm to the action  (\ref{8}), gives us the primary second class constraints 
\begin{eqnarray}\label{9}
G_{0i}&\equiv& p_{0i}-\frac12B_{ij}c_0^j=0, \cr 
G_{ni}&\equiv& p_{ni}-\pi\alpha^\prime f_nB_{ij}c_n^j=0, \cr
G_{-ni}&\equiv& p_{-ni}+\frac{f_n}{4\pi\alpha^\prime}B^{-1}_{ij}c_{-n}^j=0.
\end{eqnarray}

\noindent Their Poisson brackets are

\begin{eqnarray}\label{10}
\{G_{0i}, G_{0j}\}&=&-B_{ij}, \cr
\{G_{ni}, G_{mj}\}&=&-2\pi\alpha^\prime f_nB_{ij}\delta_{n-m,0}, \cr
\{G_{-ni}, G_{-mj}\}&=&\frac{f_n}{2\pi\alpha^\prime}B^{-1}_{ij}
\delta_{n-m,0},
\end{eqnarray}
and the corresponding Hamiltonian is
\begin{eqnarray}\label{11}
H&=&\sum_{n=1}^\infty -f_nnc_{-n}c_n +
\lambda_0(p_0-\frac 12Bc_0)+ \nonumber \\
&&\lambda_n(p_n-\pi\alpha^\prime f_nBc_n)+ 
\lambda_{-n}(p_{-n}+\frac{f_n}{4\pi\alpha^\prime}B^{-1}c_{-n})\;\;. \nonumber \\
\mbox{}
\end{eqnarray}
There are no secondary constraints in the problem. From the 
consistency conditions that constraints do not evolve in time, $\dot G=0$,
one obtains expressions for the Lagrangian multipliers 
\begin{eqnarray}\label{12}
\lambda_0=0, \quad \lambda_n=-\frac{n}{2\pi\alpha^\prime}c_{-n}B^{-1}, 
\quad \lambda_{-n}=2\pi\alpha^\prime nc_nB\;\;. \nonumber \\
\mbox{}
\end{eqnarray}
To take into account the second class constraints (\ref{9}) one 
introduces the Dirac bracket 
\begin{eqnarray}\label{13}
\{K, P\}_D=&&\{K, P\}
+\{K, G_{0i}\}(B^{-1})^{ij}\{G_{0j}, P\}+ \nonumber \cr
&&\sum_{n=1}^\infty
\{K, G_{ni}\}\frac{1}{2\pi\alpha^\prime f_n}
(B^{-1})^{ij}\{G_{nj}, P\}-\nonumber \\
&&\{K, G_{-ni}\}\frac{2\pi\alpha^\prime}{f_n}B^{ij}
\{G_{-nj}, P\}.
\end{eqnarray}
After that, the variables $p_n^i$ can be omitted from consideration. 
The resulting Hamiltonian formulation for (\ref{8}) consist of the 
physical variables $c_n^i$ with the Dirac brackets 
\begin{eqnarray}\label{14}
\{c_0^i, c_0^j\}_D&=&-(B^{-1})^{ij}, \cr
\{c_n^i, c_m^j\}_D&=&-\frac{1}{2\pi\alpha^\prime f_n}
(B^{-1})^{ij}\delta_{n-m,0}, \cr
\{c_{-n}^i, c_{-m}^j\}_D&=&\frac{2\pi\alpha^\prime}{f_n}
B^{ij}\delta_{n-m,0}, \qquad n, m>0,
\end{eqnarray}
whose dynamics is governed now by the Hamiltonian 
\begin{eqnarray}\label{15}
H=-\sum_{n=1}^\infty f_nnc^i_{-n}c^i_n.
\end{eqnarray}
As it should be for the first order system, the Hamiltonian equations 
of motion which follow from (\ref{14}), (\ref{15}) are the equations 
(\ref{7}). Note also that the variables $c_n^i$ with $n$ fixed are 
canonically conjugated to each other. 
The equations (\ref{7}) for the physical variables can be solved now 
in terms of oscillators 
\begin{eqnarray}\label{16}
c^i_n(\tau)&=&\alpha^i_ne^{in\tau}+\alpha^i_{-n}e^{-in\tau}, \cr
c^i_{-n}(\tau)&=&-2i\pi\alpha^\prime\alpha^i_nBe^{in\tau}+
2i\pi\alpha^\prime\alpha^i_{-n}Be^{-in\tau}, \cr
{\alpha^i_n}^* &=&\alpha^i_{-n}.
\end{eqnarray}
From Eq. (\ref{14}) one finds their brackets 
\begin{eqnarray}\label{17}
\{c_0^i, c_0^j\}_D&=&-(B^{-1})^{ij}, \\
\{\alpha_n^i, \alpha_m^j\}_D&=&-\frac{1}{4\pi\alpha^\prime f_{\mid n\mid}}
(B^{-1})^{ij}\delta_{n+m,0}, \quad n, m\ne 0,  \nonumber
\end{eqnarray}
while the Hamiltonian (\ref{15}) acquires the form 
\begin{eqnarray}\label{18}
H_f=4i\pi\alpha^\prime\sum_{n=1}^\infty f_nn\alpha_n^iB_{ij}\alpha_{-n}^j.
\end{eqnarray}
It is worth noting that this
procedure, being applied to the open string with the Neumann boundary
conditions, leads to the standard results \cite{gsw,polchinski}.

\section{Hamiltonian formulation for the open string with a B-field}

By using of Eqs. (\ref{4}), (\ref{16}) we restore the string coordinate 
$x^i(\tau, \sigma)$ in terms of oscillators 
\begin{eqnarray}\label{19}
x^i(\tau, \sigma)=c_0^i+\sum_{n\ne 0}\frac{e^{in\tau}}{\mid n\mid}
\left(\alpha_n^i\cos n\sigma-
2i\pi\alpha^\prime\alpha_n^jB_j{}^i\sin n\sigma\right) \nonumber \\
\mbox{} 
\end{eqnarray}
Thus, with the system (\ref{1})-(\ref{3}) one associates the 
Hamiltonian formulation which includes the variables 
$c_0^i, ~ \alpha_n^i, ~ 
n\ne 0$ with the brackets (\ref{17}) and the Hamiltonian (\ref{18}).
As a consequence of Eqs. (\ref{17}), (\ref{19})  
the induced bracket of the string coordinates turns out to be 
noncommutative in the bulk and on the boundary 
\begin{eqnarray}\label{20}
&&\lbrace x^i(\tau, \sigma), x^j(\tau, \sigma^\prime\rbrace_D \\
&=&-(B^{-1})^{ij}-
\sum_{n=1}^\infty\frac{1}{n^2f_n}\left(\frac{1}{2\pi\alpha^\prime}(B^{1})^{ij} 
\times \cos n\sigma\cos n\sigma^\prime \right. \nonumber \\
&& \left. -\,2\pi\alpha^\prime B^{ij} \times \sin n\sigma\sin n\stackrel{~}{\sigma^\prime}\right). \nonumber
\end{eqnarray}
In particular, the coordinates on D-brane obey
\begin{eqnarray}\label{21}
\{x^i(\tau, 0), x^j(\tau, 0\}_D&=& 
\{x^i(\tau, \pi), x^j(\tau, \pi\}_D  \\
&=&-\left(1+\frac{1}{2\pi\alpha^\prime}
\sum_{n=1}^\infty\frac{1}{n^2f_n}\right)(B^{-1})^{ij}\,\,. \nonumber
\end{eqnarray}
It is interesting to note that the standard normalization 
$f_n\sim\frac{1}{n}$ of the oscillator brackets (\ref{17}) implies 
that one needs to use some regularization for Eq. (\ref{21}). 
For the choice $f_n=-\frac{\pi}{12\alpha^\prime}$ the D-brane plane 
becomes commutative, and the Hamiltonian in this case is 
\begin{eqnarray}\label{22}
H=-\frac{i\pi^2}{3}\sum_{n=1}^\infty\alpha_n^iB_{ij}\alpha_{-n}^j.
\end{eqnarray}

\section{D-brane}

Finally, let us discuss freedom in definition of phase space 
bracket in the Hamiltonian formalism. Transition from configuration 
to phase space description is not unique procedure since one needs 
to define simultaneously the bracket and the Hamiltonian. 
One possibility is to start from the Poisson bracket, then exact 
expression for the Hamiltonian is known for a general case \cite{gt,de}[16, 20].
According to the Darboux theorem, one can equally take as the starting 
point any nondegenerate closed two-form and then try to define a 
Hamiltonian from the condition that corresponding equations of motion 
are equivalent to the initial ones. Turning to our case (\ref{14}), 
(\ref{15}) it will be sufficiently to consider the brackets 
\begin{eqnarray}\label{23}
\{c_0^i, c_0^j\}_D&=&-E^{ij}, \nonumber \\  
\{c_n^i, c_m^j\}_D&=&-\frac{1}{2\pi\alpha^\prime f_n}A^{ij}\delta_{n-m,0}, 
~ n, m>0,
\end{eqnarray}
with some antisymmetric nondegenerate constant matrices $E$ and $A$. 
From the condition that the dynamics (\ref{7}) is reproduced in this 
formulation one finds
\begin{eqnarray}\label{24}
\{c_{-n}^i, c_{-m}^j\}_D=\frac{2\pi\alpha^\prime}{f_n}(AB^2)^{ij}
\delta_{n-m,0}, \cr 
H=-\sum_{n=1}^\infty f_nnc_{-n}(AB)^{-1}c_n.
\end{eqnarray}
One can take Eqs. (\ref{23}), (\ref{24}) instead of Eqs. (\ref{14}), 
(\ref{15}) as the Hamiltonian formulation corresponding to the system 
(\ref{8}). Repeating the previous analysis, one finds the same 
expression (19) for the string coordinate in terms of oscillators 
which obey 
\begin{eqnarray}\label{25}
\{c_0^i, c_0^j\}_D&&=-E^{ij}, \nonumber \cr
\{\alpha_n^i, \alpha_m^j\}_D &&=-\frac{1}{8\pi\alpha^\prime f_{\mid n\mid}}
\left(A-sgn(nm)B^{-1}AB\right)^{ij} \nonumber \\
&& \times \delta_{\mid n\mid-\mid m\mid ,0},~ n, m\ne 0\,\,.
\end{eqnarray}
The quantities $f_n, ~ E, ~ A$ can be chosen in such a way 
that the noncommutativity parameter on D-brane acquires exactly the same 
form as it was obtained from the disk propagator \cite{sw}. Namely, for the 
choice 
\begin{eqnarray}\label{26}
f_n&=&\frac{\pi^2}{3(2\pi\alpha^\prime)^3}, \nonumber \\ 
E&=&\frac{(2\pi\alpha^\prime)^2}{2}A, \nonumber \\ 
A&=&\left(1+(2\pi\alpha^\prime)^2B^2\right)^{-1}B,
\end{eqnarray}
one obtains the following Hamiltonian formulation for the system 
(\ref{1})-(\ref{3})
\begin{eqnarray}\label{27}
& &H\,=\,\frac{i\pi^2}{3(2\pi\alpha^\prime)^2}\sum_{n=1}^\infty
n\alpha_nB^{-1}\left(1+(2\pi\alpha^\prime)^2B^2\right)\alpha_{-n}, \nonumber \cr
& &\{c_0^i, c_0^j\}_D\,=\,-\frac{(2\pi\alpha^\prime)^2}{2}  
\left[\left(1+(2\pi\alpha^\prime)^2B^2\right)^{-1}B\right]^{ij}, \nonumber \cr
& &\{\alpha_n^i, \alpha_m^j\}_D\,=\,-\frac{3(2\pi\alpha^\prime)^2}{2\pi^2}
\left[\left(1+(2\pi\alpha^\prime)^2B^2\right)^{-1}B\right]^{ij}\nonumber \\
&&\times \delta_{n+m,0}, ~ n, m\ne 0, 
\end{eqnarray}
which gives for the coordinates on D-brane the desired expression 
\begin{eqnarray}\label{28}
\{x^i, x^j\}_D=-(2\pi\alpha^\prime)^2(1-2i\pi\alpha^\prime B)^{-1}
B(1+2i\pi\alpha^\prime B)^{-1}. \nonumber \\
\mbox{}
\end{eqnarray}

\section{Soldering of open strings with the presence of a $B$-field}

The soldering formalism \cite{solda,aw}, is an iterative method that permits us to construct an effective action invariant under a specific gauge symmetry disclosing interesting physical features analogous to the interference phenomena.  The method uses primarily the Noether gauging procedure that helps us to fuse together the variables representing the different aspects of a theory.   It works by elevating a global symmetry to a local form thanks to a mutual cancelation of the obstruction to gauge symmetry of the individual components. It is the gauge field introduced in the process the agent responsible for this new symmetry and may be eliminated after the gauging is complete. In this Letter we will explore just the main steps of the method.  For a detailed reading, see \cite{capitulo}.

The soldering formalism is the well suited algorithm for fusing together opposite aspects of a global symmetry, such as the right and left propagating modes of chiral bosons to name just one application.  
However, the nonlinear feature of the action ({\ref{1}) can make us to face new difficulties.  
The idea is to combine two different models in order to obtain a single composite model.  The final result reflects a direct connection between the old and the new theories.  The soldering formalism accomplish this mission for theories which manifest dual aspects of some symmetry, such as chirality, self-duality \cite{bw}, noncommutativity \cite{amw,ghosh}, etc.  The final result, i.e., the soldered action, hides the mentioned symmetry in a interesting way such that it brings new informations about both theories, the old and the new ones.  It can be demonstrated that the soldering and the canonical transformation procedures have a strict relation of complementarity \cite{bg}.

In order to establish these interesting connections between the so-called parent and daughter actions \cite{ghosh} in an open string theory, let us write two different aspects of the action (\ref{1}), i.e.,
\ba
{\cal L}_{k_1}^x&=&\p_ax^i\p_ax^i+2\pi\,k_1\,\alpha^\prime\epsilon^{ab}
\partial_ax^i\partial_bx^jB_{ij} \\
{\cal L}_{k_2}^y&=&\p_ay^i\p_ay^i+2\pi\,k_2\,\alpha^\prime\epsilon^{ab}
\partial_ay^i\partial_by^jB_{ij}\,\,,
\ea
where $k_1$ and $k_2$ are connected by the relations $k_1 = k_2 = 1$ and $k_1 = - k_2 = 1$.  For simplicity we will work at this stage with the Lagrangian densities.

The relations show two open strings with equal and opposed characteristics respectively.  It is clear that the first relation is the trivial one and must show nothing new.  The relevant one must be the second relation.  However, we will carry both computations to comprove the above observations.

So, let us gauge the following global symmetry,
\be \l{6.1}
\delta x^i\,=\,\delta y^i\,=\, \sigma^i\,\,,
\ee
where $\sigma^i$ is the gauge parameter.  The gauging of the global parameter $\sigma^i$ will be done in the soldering process.  With this symmetry, a simple calculation shows that,
\be \l{6.2}
\delta {\cal L}_0^{k_1}\,=\, J_{xj}^{b}\,\p_b\,\sigma^j\,\,,
\ee
where $J_x^{bj}$ is Noether chiral current given by
\be \l{6.3}
J_{xj}^{b}\,=\,2[\p^b\,x_j\,+\,2\,\pi\,\alpha'\,k_1\,B_{ij}\epsilon^{ab}\p_a\,x^i]
\ee
that parameterizes the lack of gauge symmetry of the original action ${\cal L}_0^x$.   

The next step is to introduce an auxiliary field that helps in the gauging procedure.  Let us call this field by $D^{bi}$, known as the soldering field.  We can now construct the first-iterated correction of the open string action (\ref{1}) as 
\be \l{6.4}
{\cal L}_0^{k_1} \rightarrow {\cal L}_1^{k_1}\,=\,{\cal L}_0^{k_1}\,-\,D^i_b\,J^{b}_{xi}\,\,.
\ee
Notice that we are looking for an action which is gauge-invariant under (\ref{6.1}).

The gauge variation of ${\cal L}_1^x$ is,
\ba \l{6.5}
\delta {\cal L}_1^{k_1}&=&-D^i_b\,\delta J_{xi}^{b} \nonumber \\
&=&-\delta (D^j_b)^2\,-\,4\,\pi\,\alpha'\,k_1\,B_{ij}\,\epsilon^{ab}\,(\delta D^i_a)D^j_b
\ea
where we have chosen $\delta D^i_b=\p_b \sigma^i$ to cancel (\ref{6.2}) and used (\ref{6.1}) once again.

Again we can construct the second-iterated correction of the open string action (\ref{1}) as,
\be \l{6.6}
{\cal L}_1^{k_1} \rightarrow {\cal L}_2^{k_1}\,=\,{\cal L}_0^{k_1}\,+\,(D^j_b)^2
\ee
and therefore,
\be \l{6.7}
\delta {\cal L}_2^{k_1}\,=\,-4\pi\alpha'\,k_1\,B_{ij}\epsilon^{ab}(\delta D^i_a)\, D^j_b
\ee

Accomplishing the same procedure for ${\cal L}_y^{k_2}$ we have that,
\be \l{6.8}
\delta {\cal L}_2^{k_2}\,=\,-4\pi\alpha'\,k_2\,B_{ij}\epsilon^{ab} D^i_a\,(\delta D^j_b)
\ee

Because of the antisymmetric characteristic of both tensors, $B_{ij}$ and $\epsilon_{ij}$ it can be easily demonstrated that, for $k_1=k_2=1$ and $k_1=-k_2=1$ we have that,
\ba \l{6.9}
{\cal L}^{k_1=k_2=1}_{final} &=& {\cal L}_2^{k_1}\,+\,{\cal L}_2^{k_2}\,+\,4\pi\alpha'\,B_{ij}\epsilon^{ab}\,D^i_a\,D^j_b \nonumber \\
{\cal L}^{k_1=-k_2=1}_{final} &=& {\cal L}_2^{k_1}\,+\,{\cal L}_2^{k_2}\,\,.
\ea
In both cases we have
\be \l{6.10}
\delta {\cal L}_2\,=\,0\,\,,
\ee
to finally obtain the desired gauge-invariant action.

In order to be brief, the next step is to eliminate the auxiliary field $D^i_a$ through its equations of motion in each final action above separately, which demands a certain algebra.  After that we have to substitute it back in both actions (\ref{6.9}) and finally obtain two effective actions, respectively,
\ba \l{6.11}
&& {\cal S}^{k_1=k_2=1}_{eff} \\
&=&-\frac{1}{4\pi\alpha^\prime}\int d^2\sigma [{1\over2}\,\p^a z^i\,\p^a z^i
\,+\,\pi\,\alpha'\,B_{ij}\epsilon^{ab}\,\p_a z^i\,\p_b z^j]\,\,, \nonumber
\ea
and
\be \l{6.12}
{\cal S}^{k_1=-k_2=1}_{eff} \,=\,-\frac{1}{4\pi\alpha^\prime}\int d^2\sigma [{1\over2}\,\p^a z^i\,\p^a z^i]\,\,,
\ee
where in both cases we have a compose string coordinate given by $z^i=x^i-y^i$.  

In the first case, the soldered action (\ref{6.11}) shows us that the fusion (soldering) of two open strings with the aspects results in another open string.  However, we have a different behavior when both strings have opposite aspects.  The result (\ref{6.12}) shows a Polyakov action which is well known to be equivalent at the classical level to the Nambu-Goto action.  This last result establish a direct connection between the parent and the daughter action, i.e., between the open string action and the Polyakov action.  Moreover, we can see clearly that in the first case, the soldering coupling keeps the magnetic field $B_{ij}$.  However, in the second case the magnetic field disappear from the effective action.\\


\noindent {\bf The dual projection analysis:}  Let us analyze the Polyakov action given by
\be \l{6.13}
{\cal S}\,=\,-\frac{1}{4\pi\alpha^\prime}\int d^2\sigma \sqrt{g}g^{ab}\p_a x_{\mu}\p_b x^{\mu}\,\,,
\ee
which is invariant under manifest reparametrization invariance
\ba \l{6.14}
\delta x^{\mu}&=&\epsilon^a\,\p_a x^{\mu} \nonumber \\
\delta g^{ab}&=&\epsilon^c \p_c g^{ab}\,-\,b^{ac} \p_c\epsilon^b\,-\,g^{bc}\p_c \epsilon^a \nonumber \\
\delta \sqrt{g}&=&\p_a (\epsilon^a \sqrt{g})\,\,.
\ea
We can easily notice that the Polyakov action above resembles an action with scalar fields interacting with an external two-dimensional gravitational field.

Recently, some of us showed that the dual projection mechanism \cite{aw,eu} disclosed from the scalar model embbeded in a gravitational background, two particles with no dynamics, well known as notons.  The noton field was introduced by Hull in order to cancel off the Siegel anomaly \cite{hull}.

Having in mind this idea, we can assume that the Polyakov action has a relation with the noton as,
\ba \l{6.15}
{\cal S} \,=\,-\frac{1}{4\pi\alpha^\prime}\int d^2\sigma & &[\dot{x}_{+\mu} {x'}^{\mu}_+  \,-\,\eta_+ ({x'}^{\mu}_+)^2 \nonumber \\
& & \,-\,\dot{x}_{-\mu} {x'}^{\mu}_-\,-\,\eta_- ({x'}^{\mu}_-)^2]
\ea
where $\eta_{\pm}$ is a function of the metric elements and $x_{\pm}^{\mu}$ are the notons (for more details see \cite{aw,eu}).   So, we can see in (\ref{6.15}) that the Polyakov action can be perceived as a combination of two notons.
As the soldering formalism establish a relation between the parent and the daughter actions we can say from (\ref{6.11}) that there is a relation between the noton field and the open string action.

\section{conclusions}

In this work we quantized the open string action used by Seiberg and Witten following precisely the standard methods.  We obtained the equations of motion for the mechanical variables and wrote the respective action.  The mechanical variables are the ones that can be taken in order to generate the string coordinate.  The equations of motion permit us to introduce the Lagrangian multipliers.

The direct application of the Dirac formalism gives us the primary second-class constraints and consequently the Poison brackets and the corresponding Hamiltonian.  Since there is no secondary constraints in the problem we can construct the Dirac brackets for the theory and the resulting Hamiltonian with the physical variables.  We also constructed a Hamiltonian for the physical variables written in terms of oscillators.  With this relation in terms of oscillators we restored the string coordinate and computed the Dirac brackets for these coordinates.

Writing the Dirac brackets in terms of antisymmetric nondegenerate constant matrices we obtained the Hamiltonian formulation for the original system (1)-(3).  These matrices can be chosen in such a way that the noncommutativity parameter on $D$-brane acquires precisely the same form as it was obtained in \cite{sw}.

Finally, the soldering formalism permit us to establish a clear relation between the original system and the Polyakov action which is equivalent to the Nambu-Goto theory.  The Polyakov action clearly resembles an action with scalar fields interacting with an external two-dimensional gravitational field.  This relation bring us the idea that there must be a kind of noton particle in the Polyakov formulation since it was demonstrated by two of us that the noton comprise the spectrum of scalar fields with gravitational background.

\section{ Acknowledgments}

The authors would like  to thank CNPq (Conselho Nacional de Desenvolvimento Cient\' ifico e Tecnol\'ogico) and FAPEMIG (Brazilian financial agencies) for financial support.


\begin{thebibliography}{nn}

\bibitem{cds} A. Connes, M. R. Douglas and A. Schwarz, JHEP 02 (1998) 003, hep-th/9711162.

\bibitem{sw} N. Seiberg and E. Witten, JHEP 09 (1999) 032, hep-th/9908142.

\bibitem{clny} C. G. Callan, C. Lovelace, C. R. Nappi and S. A. Yost, Nucl. Phys. B 288 (1987) 525; 
A. A. Abouelsaood, C. G. Callan, C. R. Nappi and S. A. Yost, Nucl. Phys. B 280 (1987) 599.

\bibitem{as} H. Arfaei and M. M. Sheikh-Jabbari, Nucl. Phys. B 526 (1998) 278. 

\bibitem{sheikh} M. M. Sheikh-Jabbari,  Phys. Lett. B 425 (1998) 48.
  
\bibitem{ft} E. S. Fradkin and A. A. Tseytlin, Phys. Lett. B 163 (1985) 123.

\bibitem{schomerus} V. Schomerus, JHEP 06 (1999) 030, hep-th/9903205.

\bibitem{aas}   F. Ardalan, H. Arfaei and M. M. Sheikh-Jabbari, {\em Mixed Branes and M(atrix) Theory on Noncommutative Torus}, hep-th/9803067. 

\bibitem{aas2}   F. Ardalan, H. Arfaei and M. M. Sheikh-Jabbari  JHEP 02 (1999) 016, hep-th/9810072.

\bibitem{aas3}   F. Ardalan, H. Arfaei and M. M. Sheikh-Jabbari   Nucl. Phys. B 576 (2000) 578, hep-th/9906161.

\bibitem{ch}   C. S. Chu and P. M. Ho, Nucl. Phys. B 550 (1999) 151, hep-th/9812219.

\bibitem{ch2}   C. S. Chu and P. M. Ho, Nucl. Phys. B 568 (2000) 447, hep-th/9906192.

\bibitem{gj} S. Girvin and T. Jach, Phys. Rev. D 29 (1984) 5617. 

\bibitem{dj} G. Dunne and R. Jackiw, Nucl. Phys. Proc. Suppl. 33 C (1993) 114, hep-th/9204057. 

\bibitem{bs} D. Bigatti and L. Susskind, Phys. Rev. D 62 (2000) 066004, hep-th/9908056.

\bibitem{lee} T. Lee, Phys. Rev. D 62 (2000) 024022, hep-th/9911140.

\bibitem{ko} W. T. Kim and J. J. Oh, Mod. Phys. Lett. A 15 (2000) 1597, hep-th/9911085.

\bibitem{dirac} P. A. M. Dirac, {\em Lectures on Quantum Mechanics}, Yeshiva University, NY, 1964.

\bibitem{gt} D. M. Gitman and I. V. Tyutin, {\em Quantization of Fields
with Constraints}, Springer-Verlag, 1990.

\bibitem{bh} L. Brink and M. Henneaux, {\it{Principles of String Theory}}, Plenum Press, NY and London, 1988.

\bibitem{gsw} M.B. Green, J.H. Schwarz and E. Witten, {\it{Superstring Theory}}, v.1,2, Cambridge Univ. Press, Cambridge, 1987.

\bibitem{polchinski} J. Polchinski, {\it{String Theory}}, v.1,2, Cambridge University Press, 1998.

\bibitem{de} A. A. Deriglazov and K. E. Evdokimov, Int. J. Mod. Phys. A 15 (2000) 4045, hep-th/9912179.

\bibitem{solda} M. Stone, Phys. Rev. Lett. 63 (1989) 731; Nucl. Phys. B 327
(1989) 399; Report no. ILL-23/89 (unpublished); D. Depireux, S. J. Gates Jr.
and Q-Han Park, Phys. Lett. B 224 (1989); E. Witten, Commun. Math. Phys. 144
(1992) 189; E. M. C. Abreu, R. Banerjee and C. Wotzasek, Nucl. Phys. B 509
(1998) 519; R. Amorim, A. Das and C. Wotzasek, Phys. Rev. D 53 (1996) 5810; 

\bibitem{aw}   E. M. C. Abreu and C. Wotzasek, Phys. Rev. D 58, 101701(R) (1998).

\bibitem{ghosh}   S. Ghosh, Phys. Lett. B 579 (2004) 377.

\bibitem{capitulo}  C. Wotzasek, ``Soldering Formalism: theory and applications", hep-th/9806005;
E. M. C. Abreu and C. Wotzasek, ``Topics on the Quantum Dynamics of Chiral Bosons", in ``Progress in Boson Research" Nova Eds., NY, USA, pp. 259-318, hep-th/0410019.

\bibitem{bw}  R. Banerjee and C. Wotzasek, Nucl. Phys. B 527 (1998) 402.

\bibitem{bg}  R. Banerjee and S. Ghosh, Phys. Lett. B 482 (2000) 302.

\bibitem{amw}  E. M. C. Abreu, R. Menezes and C. Wotzasek, Phys. Rev. D 71 (2005) 065004.

\bibitem{hull}   C. M. Hull, Phys. Lett. B 206 (1988) 234; B212 (1988) 437.

\bibitem{eu}   E. M. C. Abreu, Int. J. Mod. Phys. A 19 (2004) 1823.
  


 

\end{thebibliography}
\end{document}